\documentclass[aps,pre,twocolumn,amsmath,amssymb,superscriptaddress,showpacs,citeautoscript]{revtex4}
\usepackage{graphicx}
\usepackage{color}
\renewcommand{\vec}[1]{\mathbf{#1}}

\begin{document}
\title{Crossovers in the dynamics of supercooled liquids probed by an amorphous wall}

%\date{\today}

\author{Glen M. Hocky}
\affiliation{Department of Chemistry, Columbia University, 
3000 Broadway, New York, New York 10027, USA}

\author{Ludovic Berthier}
\author{Walter Kob}
\affiliation{Laboratoire Charles Coulomb, UMR 5221, CNRS and Universit\'e
Montpellier 2, 34095 Montpellier, France}

\author{David R. Reichman}
\affiliation{Department of Chemistry, Columbia University, 
3000 Broadway, New York, New York 10027, USA}

\begin{abstract}
We study the relaxation dynamics of a binary Lennard-Jones liquid in
the presence of an amorphous wall generated from equilibrium particle
configurations. In qualitative agreement with the results presented in
Nature Phys. {\bf 8}, 164 (2012) for a liquid of harmonic spheres, we find
that our binary mixture shows a saturation of the dynamical length scale
close to the mode-coupling temperature $T_c$.
Furthermore we show that, due to the broken symmetry imposed by the wall, 
signatures of an additional
change in dynamics become apparent at a temperature well above
$T_c$. We provide evidence that this modification in the relaxation dynamics
occurs at a recently proposed dynamical crossover temperature $T_s >
T_c$, which is related to the breakdown of the Stokes-Einstein relation.
We find that this dynamical crossover at $T_s$ is also observed for the
harmonic spheres as well as a WCA liquid, showing that it
may be a general feature of glass-forming systems.

\end{abstract}

\pacs{64.70.Q-,05.20.Jj,05.10.-a}
%05.10.-a 	Computational methods in statistical physics and nonlinear dynamics (see also 02.70.-c in mathematical methods in physics )
%05.20.Jj 	Statistical mechanics of classical fluids (see also 47.10.-g General theory in fluid dynamics)
%64.70.Q- 	Theory and modeling of the glass transition 

\maketitle

\section{Introduction}
\label{sec:intro}
There has been a great surge in studies which
seek to investigate length scales in supercooled
liquids through simulation in the presence of quenched amorphous order
\cite{Biroli-NatPhys2008,Kim-EPL2009,Kob-NatPhys2012,Jack-PRE2012,karmakar-pinned,Berthier-static-PRE2012,Charbonneau-PRL2012,Hocky-PRL2012,Charbonneau-PRE2013,Kob-PRL2013,Jack-PRE2013}.
This recent interest has been spurred on by the  successful
computational realization of thought experiments predicting growing
static length scales on increased supercooling, as well as the
ever-growing availability of CPU time which makes such studies feasible
\cite{Bouchaud-JCP2004,Biroli-NatPhys2008,Cammarota-PRL2011,Berthier-static-PRE2012,Charbonneau-PRL2012,Hocky-PRL2012,Charbonneau-PRE2013,Kob-PRL2013,karmakar-pinned,Jack-PRE2013}.

Yet, studies seeking to understand growing dynamical length
scales by employing simulations with frozen particles go back
even earlier \cite{Scheidler-EPL2002,Kim-EPL2003}. The authors of
Ref.~\onlinecite{Scheidler-EPL2002}, in particular, studied the Kob-Andersen
binary Lennard-Jones system (KA) in the presence of a rough wall
created by fixing the positions of a slab of particles from a bulk
equilibrium configuration. They found that the dynamics near the
rough wall slowed down substantially and the analysis of the profiles
of relaxation times near the wall provided a length scale which grew
with decreasing temperature. A recent study revisited these ideas in a
supercooled harmonic sphere system (HARM) for larger sample sizes and
down to very low temperatures, below the mode-coupling temperature $T_c$
\cite{Kob-NatPhys2012}. This study revealed a dynamical length scale that
first increased as $T$ approached $T_c$ from above, and then surprisingly
decreased for $T<T_c$ \cite{Kob-NatPhys2012}. This behavior was attributed
to a change in the dynamics below $T_c$ where collective particle
rearrangements become predominant \cite{Kob-NatPhys2012,Kob-PP2012,Kob-arxiv2014}. Such
a change is naturally understood in the framework of the random first
order transition (RFOT) theory \cite{Stevenson-NatPhys2006,Berthier-RMP2011},
where a crossover between non-activated correlated relaxation to thermally
activated cooperative dynamics is expected to take place. 

It should be noted
that the dynamical length scale defined in Ref.~\onlinecite{Kob-NatPhys2012} need not 
correspond to that extracted from study of the bulk four-point correlation function $S_4(k,t)$,
but simply defines an independent dynamic correlation length scale associated with the 
spatial extent of the perturbation of dynamic relaxation near an amorphous wall.
Although there is a formal connection between bulk correlations and dynamic response to 
an infinitesimal field in the linear response regime \cite{Berthier-JCP2007,Kim-JPCB2013}, 
the connection between these different approaches to measure dynamic correlation lengthscales 
in systems with glassy dynamics will not be addressed in this work, but is a worthy topic 
for future research.

Since at present the non-monotonic $T-$dependence of the dynamical
length scale probed by an amorphous wall
has been observed only in one model glass-former,
it is important to investigate whether this phenomenon is general
or not.  In order to address this question we have performed a similar
analysis in the widely studied KA system (the same model used in
Ref.~\onlinecite{Scheidler-EPL2002}).  When taken together, the results of
previous studies in Refs.~\onlinecite{Berthier-finite-PRE2012,Berthier-JCP2007,Flenner-JCP2013} 
may be interpreted as suggesting that a crossover in the relaxation
dynamics across the mode-coupling temperature exists but is weaker
for the KA than the HARM system. Therefore we anticipate that if the
non-monotonic evolution of dynamic profiles near an amorphous wall results
from this crossover, then this effect should be less pronounced in the
KA model. In the present work, we seek to assess this possibility and to
better understand which features of supercooled liquids near an amorphous
wall are generic. Given our findings in the KA system, we then extend
our study to two other supercooled liquids. Furthermore we demonstrate
that indications of an additional crossover at a temperature higher
than $T_c$ may be uncovered by the investigation of dynamics close to
an amorphous wall, therefore showing that there are in fact {\it two}
crossover temperatures at which the dynamics is changing.

This paper is organized as follows. In Sec.~\ref{sec:methods} we
discuss the models to be studied, as well as the details of the various
calculations we will employ. In Sec.~\ref{sec:overlap} we present the
results of our analyses and comparison with the previous results of
Ref.~\onlinecite{Kob-NatPhys2012}. In Sec.~\ref{sec:perppar} we present
results comparing the dynamics in directions perpendicular and parallel
to the wall, which reveal further information related to the dynamics of
supercooled liquids. Finally, in Sec.~\ref{sec:conclusions} we conclude
and discuss the impact of our results in the broader context of recent
work on supercooled liquids.

\section{Models and Methods}
\label{sec:methods}

In this paper, we present data for three model systems. Our primary system
of interest is the Kob-Andersen Lennard-Jones model (KA), an 80:20 binary
mixture of particles at density $\rho=1.2$ with very well characterized
structural and dynamical properties \cite{Kob-PRE1995a,Kob-JPCM1999}. All quantities
are reported in standard reduced units. In this model, we find the onset of slow
dynamics occurs near $T_o \approx 1.0$, and the mode-coupling crossover
has been previously reported as $T_c \approx 0.435$ \cite{berthier-tarjus}. 
We have fully equilibrated this model for the
system size $N=1900$ particles in a cubic box for the temperatures $T=0.9$,
0.8, 0.7, 0.65, 0.625, 0.6, 0.575, 0.56, 0.55, 0.5, 0.48, 0.45, 0.435, and 0.432.
Simulations were done with $NVT$ dynamics for $100\tau_\alpha^{\rm bulk}$
at each temperature using the LAMMPS package \cite{Plimpton-JCP1995};
$\tau_\alpha^{\rm bulk}$ was determined by calculating the self-intermediate
scattering function, $F_s(k,t)=\frac{1}{N}\sum_{j=1}^N e^{i
\vec{k}\cdot(\vec{x}_j(t)-\vec{x}_j(0))}$, and defining $F_s(k=7.25,t
\equiv \tau_\alpha^{\rm bulk})=1/e$.  Each equilibrated configuration was
replicated three times along the $z-$axis to make a rectangular box
of dimensions approximately $11.655 \times 11.655 \times 34.966$. The
rectangular boxes were again simulated for $100\tau_\alpha^{\rm bulk}$
to remove the periodicity introduced by replicating the system.  Each
configuration was then tested for equilibration by calculating $F_s(k,t)$
for the first and second half of yet another $100\tau_\alpha^{\rm bulk}$
length trajectory. Only configurations whose dynamics showed no signs
of aging (i.e. identical scattering functions for both halves of the
trajectory) were used for subsequent steps.

To study these configurations with an amorphous wall, simulations were
run where the positions of particles within a slab of width $W=3$ were
held fixed. Since we simulate the KA system with a standard cutoff of
the potential at $r_{\rm cut}=2.5$, this slab is effectively of infinite
thickness as no particle on one side can interact directly or indirectly
with those on the other side. This allows to determine the $z-$dependence
of the static and dynamic properties of the system for distances up to
$z_{max}=(34.966-3)/2\approx15.98$. At each temperature, we ran molecular
dynamics simulations of length $500$-$1000\tau_{\alpha}^{\rm bulk}$ for
$30-45$ independent wall realizations to ensure both thermalization of
a single realization, and a proper disorder average over the quenched
disorder imposed by the frozen wall.

In addition to the KA system, we performed a similar set of studies for
two other models. The first is the Weeks-Chandler-Andersen (WCA) version
of the KA system \cite{Weeks-JCP1971}, for which we performed simulations
down to $T=0.325$ (the mode-coupling temperature is $T_c \approx 0.28$
\cite{berthier-tarjus}, and $T_o\approx0.7$).
For this model, identical box sizes were
used as for the KA. The third model we study is the harmonic
sphere system (HARM) of Ref.~\onlinecite{Kob-NatPhys2012}, which had been 
previously discussed \cite{OHern-PRL2002,berthier-witten} ($T_c\approx5.2$ \cite{Kob-NatPhys2012},
 $T_o\approx 12$). As
in Ref.~\onlinecite{Kob-NatPhys2012}, we use $N=4320$, however here we
prepare the system analogously to the KA system, first equilibrating
a sample containing 1440 particles, then replicating it to construct 
a box with dimensions
approximately $12.873\times12.873\times38.620$. In this case, we used
a wave vector $k=6.28$ to define the relaxation time of the system and
other subsequent quantities, and a wall thickness of $W=2$, due to the
very short ranged potential in this model. For the HARM and WCA models
we ran simulations with $20-30$ independent wall realizations. For the
WCA system we found, by looking at aging behavior in $F_s(k,t)$, that some
trajectories crystallized after times on the order of several hundred bulk
$\tau_{\alpha}$ at $T=0.35$ in the presence of the wall. The tendency to
crystallize was more pronounced at $T=0.325$. Crystallization was evident
from visual inspection as well as from a drop in the potential energy
of samples which crystallized. These trajectories were excluded from
our data, and hence our statistical confidence at these temperatures is
reduced and a detailed analysis of temperatures close to the mode-coupling
crossover was not possible.

For each model and for each trajectory, we have calculated the overlap,
$q_c(z,t)$, and self-overlap, $q_s(z,t)$, as a function of time and
distance ($z$) from the face of the wall. These quantities, which
contain similar information as the coherent and incoherent intermediate
scattering functions, are calculated by tiling the system outside of
the wall into small boxes of side-length $l$ small enough such that
they have occupation numbers (defined by the number of particle centers
in a cell) $n_j\in\{0,1\}$. In a previous work it was determined that
$l\approx0.37$ was a good size to enforce this condition in the KA and
WCA systems without making $l$ so small as to result in poor statistics
\cite{Hocky-PRL2012}. In the present study we choose $l = 0.37597$ which
tiles each short dimension of the samples into $\tilde{N}=31$ boxes, but
we have checked that changing this parameter does not have any effect
on the resulting physics. Similarly, we choose $l\approx0.45$ for the
HARM, and tested that this gives the same results as $l\approx0.55$,
the value used in Ref.~\onlinecite{Kob-NatPhys2012} (at that size, $n_j>1$
occasionally). We define, at each distance $z$ from the wall,
\begin{equation}
q_c(z,t) = \left \langle \frac{1}{\tilde{N}^2} 
\sum_{j=1}^{\tilde{N}^2}\frac{n_j(t'+t)n_j(t')}{\rho l^3} \right \rangle,
\label{eq:overlap}
\end{equation}
\noindent
and
\begin{equation}
q_s(z,t) = \left \langle \frac{1}{\tilde{N}^2}
\sum_{j=1}^{\tilde{N}^2}\frac{p_j(t)}{\rho l^3} \right \rangle,
\label{eq:self-overlap}
\end{equation}
\noindent
where $p_j(t)$ is a function which is 1 if box $j$ is occupied by a particle at 
a reference time  $t'$ and later by the same particle at $t'+t$. In a
bulk system with no wall, $n_j(t)$ and $n_j(0)$ fully decorrelate as $t
\to \infty$, so the correlation function $q_c(z,t)$ tends to the random
value $q_r \equiv \rho l^3$. Hence we define $\tilde{q}_c(z,t)\equiv
q_c(z,t)-q_r$ so that this function decays to zero in the bulk case. With
the presence of the wall, this function will decay to a finite value
termed $\tilde{q}_c(z,\infty)\geq 0$, which quantifies the strength of
the static correlations imposed by the presence of the wall. We define the
relaxation time of the self-overlap $\tau_s(z)$ by $q_s(z,\tau_s(z))=0.2$
\footnote{Ref.~\onlinecite{Kob-NatPhys2012} defined the relaxation time
of the self-overlap $\tau_s$  by fitting the final decay of $q_s(z,t)$
to a stretched exponential of the form $A \exp[(-t/\tau_s)^\eta]$ with
$\eta<1$. We found the definition in the main text to give relaxation
times proportional to these, but due to the increased computational
expense of simulating the KA system, we chose this definition, which
allows us to extract relaxation times at distances near the wall at
the lowest temperatures with shorter simulations than would otherwise
be necessary.}.

We have also investigated a second measure of density relaxation, the
self-intermediate scattering of particles found at distance $z$ from
the wall,
\begin{equation}
F_s(\vec{k},z,t) = \left \langle \frac{1}{N(z)} \sum_{j=1}^{N(z)} 
e^{i \vec{k}\cdot(\vec{x}_j(t)-\vec{x}_j(0))} \right \rangle,
\label{eq:fsk}
\end{equation}
\noindent
where this sum extends over the particles which start in a slab of width 1.0 at distance $z$
from the wall, but need not necessarily end in
that slab. We then define scattering functions for relaxation parallel
and perpendicular to the wall by using wave vectors which are aligned with
the unit vectors $\hat{\vec{x}}$ and $\hat{\vec{z}}$ respectively. Thus,
we define $F_s^{||}(z,t) \equiv F_s(\vec{k}=7.25\hat{\vec{x}},z,t)$
and $F_s^{\perp}(z,t) \equiv F_s(\vec{k}=7.25\hat{\vec{z}},z,t)$, and
define $\tau^{||}$ and $\tau^{\perp}$ when the associated correlation
functions decay to a value of $1/e$.

\section{Overlap relaxation times}
\label{sec:overlap}

\subsection{Saturating dynamic length scale}

The first question we wish to address is whether the non-monotonic growth
of the dynamical length scale seen in Ref.~\onlinecite{Kob-NatPhys2012}
is also found in models other than the HARM system. For this purpose
we measure dynamical relaxation profiles by calculating $\tau_s(z)$ and
$\tau_c(z)$ at a series of temperatures in the KA system. As discussed in
Ref.~\onlinecite{Kob-NatPhys2012}, dynamical properties calculated from
the self and collective overlaps give very similar physical results. However,
as $q_c(z,t)$ decays to a plateau while $q_s(z,t)$ always decays to
zero, it is easier to measure $\tau_s(z)$ with confidence as there is
no need to include the long time limit of this correlation function as
an additional fitting parameter. We therefore concentrate on properties
derived from the self-overlap $q_s(z,t)$.

Before discussing these properties we note that, just as in the HARM
system, the static overlap $\tilde{q}_c(z,\infty)$ decays exponentially
with the distance $z$ at all temperatures in the KA system, and that
the static length extracted by fitting this decay grows by only a small
fraction of its value at high temperatures over the considered temperature
range. These properties are discussed in Appendix \ref{sec:static}.

\begin{figure}[t]
\centering
\includegraphics{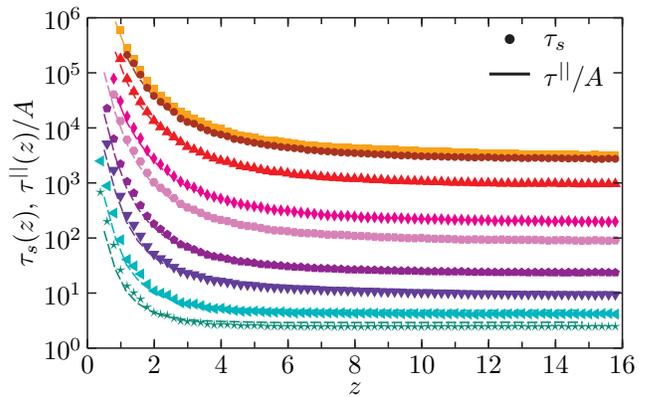}
\caption{(Color online) Relaxation times in the KA system from self-overlap ($\tau_s$, points)
and parallel self-intermediate scattering function ($\tau^{||}$, lines) versus the
distance $z$ from the wall, shown for all studied temperatures. Error bars
as determined from a bootstrap analysis \cite{Efron1993-Bootstrap} are much smaller than the points shown. The
second set of relaxation times is rescaled on the first one using a
temperature independent rescaling constant $A=0.7$ obtained from $A=\tau^{||}(z=15,T=0.435)/\tau_s(z=15,T=0.435)$. From top to bottom: $T= 0.432, $
0.435, 0.45, 0.48, 0.5, 0.55, 0.6, 0.7, and 0.8. }
\label{fig:tau_comparison}
\end{figure}

The relaxation times $\tau_s(z)$ for different temperatures are shown
in Fig.~\ref{fig:tau_comparison}. As expected from previous studies,
the dynamics slow dramatically near the wall, and relaxation
times at a given distance increase with decreasing temperature
\cite{Scheidler-EPL2002,Kob-NatPhys2012}. At large $z$ each curve tends
to plateau before $z$ reaches a value 
corresponding to half of the box length. For
comparison, we show that the relaxation times $\tau^{||}$ scale directly on top of the data
from the self-overlap.
demonstrating that the relaxation of the self-overlap is mostly dominated by
relaxation in the parallel direction to the wall. This is reasonable since
the dynamics in the direction perpendicular to the wall is significantly
slower (see Sec.~\ref{sec:perppar} below), so that the self-overlap has
already decayed by the time perpendicular motion sets in.

Based on earlier results for the KA model and results in the HARM
system, we make the ansatz that the logarithm of the relaxation
times in the presence of the wall decays exponentially to a plateau
\cite{Scheidler-EPL2002,Kob-NatPhys2012},
\begin{equation}
\ln(\tau_s(z)) = B \exp \left( - z/\xi^{\rm exp}_{\rm dyn} \right) + \ln(\tau_s^b),
\label{eq:tauexp}
\end{equation} 
where $\tau_s^b$ is the bulk relaxation time, i.e. the relaxation time computed
in the same manner but without the presence of the wall.

\begin{figure}[t]
\centering
\includegraphics{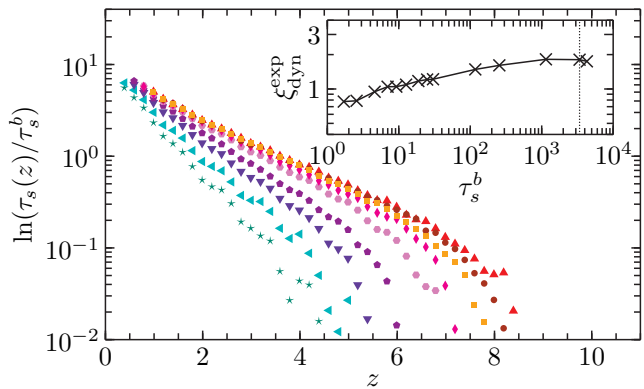}
\caption{(Color online) Self-overlap relaxation times in the KA system at a distance $z$
from the wall are divided by the bulk relaxation for the given temperature. The temperatures and symbols shown are the same as in Fig.~\ref{fig:tau_comparison}. By fitting these curves to
Eq.~(\ref{eq:tauexp}), a dynamical length scale $\xi_{\rm dyn}^{\rm exp}$
is extracted, as shown in the inset for all temperatures listed in Sec.~\ref{sec:methods}.
This length scale grows with
decreasing temperature but saturates at a temperature above $T_c$,
where $\tau_s^b \approx 3 \times 10^3$, indicated by a vertical dotted line. }
\label{fig:taus_scaled_exponential}
\end{figure}

To extract the temperature dependence of the lengths
in Fig.~\ref{fig:tau_comparison} we follow the recipe of
Ref.~\onlinecite{Kob-NatPhys2012}. At each temperature we divide the data 
by $\tau_s^b(T)$ (these data can be found in Appendix \ref{sec:tauT}).  We then fit the decay of $\ln(\tau_s(z)/\tau_s^b)$
to Eq.~(\ref{eq:tauexp}) and extract the length scale $\xi^{\rm exp}_{\rm dyn}$
and the prefactor $B$. The results of this procedure can be seen in
Fig.~\ref{fig:taus_scaled_exponential}.

The inset of Fig.~\ref{fig:taus_scaled_exponential} shows the growth of
the obtained dynamical length scale $\xi^{\rm exp}_{\rm dyn}$ and we recognize
that $\xi^{\rm exp}_{\rm dyn}$ saturates for $T>T_c$. That this saturation is not
just an artifact of the fitting procedure can be easily recognized from
the data shown in the main panel of Fig.~\ref{fig:taus_scaled_exponential}
which, for low temperatures, collapses almost perfectly onto a single
master curve which implies that $\xi^{\rm exp}_{\rm dyn}$ depends only weakly on
$T$. From this saturation we can conclude that our observed $T-$dependence
of $\xi^{\rm exp}_{\rm dyn}$ is compatible with the results obtained for the
HARM system~\cite{Kob-NatPhys2012}, even if for the  KA system there
is no strong evidence for a non-monotonic behavior in $\xi^{\rm exp}_{\rm dyn}$
(although the length does appear to decrease very slightly). The
absence of such a non-monotonic $T-$dependence in the KA system 
does not preclude the possibility of a change in behavior of 
$\xi_{\rm dyn}^{\rm exp}$ at lower temperatures. Indeed, it is expected that 
at temperatures below $T_c$, $\xi_{\rm dyn}^{\rm exp}$ will increase again at 
a temperature that is system dependent, and this behavior could supersede the
non-monotonicity observed in the length scale of the HARM system \cite{Kob-PP2012}. 

That the $T-$dependence of $\xi^{\rm exp}_{\rm dyn}$ is less pronounced for
the KA model than the HARM is also consistent with the results
from Ref.~\onlinecite{Berthier-finite-PRE2012} in which the relaxation
dynamics of both KA and HARM has been studied using periodic boundary
conditions. In that work it was shown that finite size effects can lead
to a non-monotonic $T-$dependence of the relaxation times for the HARM
system, but that in the KA system these effects are much less pronounced,
a result that was argued to be related to how sharp the cross-over between
mode-coupling like dynamics to activated dynamics is (less sharp for
the KA system than for the HARM model), which is reflected directly
in the $T-$dependence of the dynamical length scale (less pronounced
for the KA than for the HARM). This difference between the two systems
is also in agreement with the results from Ref.~\onlinecite{Flenner-JCP2013}
where it was shown that for the HARM system the height of the peak
in the dynamical four-point correlation function $\chi_4^{NVE}$ has  a
non-monotonic behavior in $T$ whereas the one for the KA system shows
only a saturation~\cite{Berthier-JCP2007}. Finally we mention that
the existence of a non-monotonic behavior in $\xi^{\rm exp}_{\rm dyn}$, or its
saturation, can be naturally interpreted in the context of RFOT
in terms of an underlying change of physical
mechanism responsible for structural relaxation occurring at $T_c$
\cite{Kob-NatPhys2012,Stevenson-NatPhys2006,Kob-PP2012,Kob-arxiv2014}.

\subsection{Further analysis of dynamic profiles}

While the decay of $\ln(\tau_s(z))$ near the wall does indeed appear
to match the exponential ansatz well, we observe that the relaxation times $\tau_s^b$
are larger than the relaxation time at $z=15.98$ (see Appendix \ref{sec:tauT}).
This suggests that the analysis of the dynamic profiles is in fact not totally straightforward.

\begin{figure}
\centering
\includegraphics{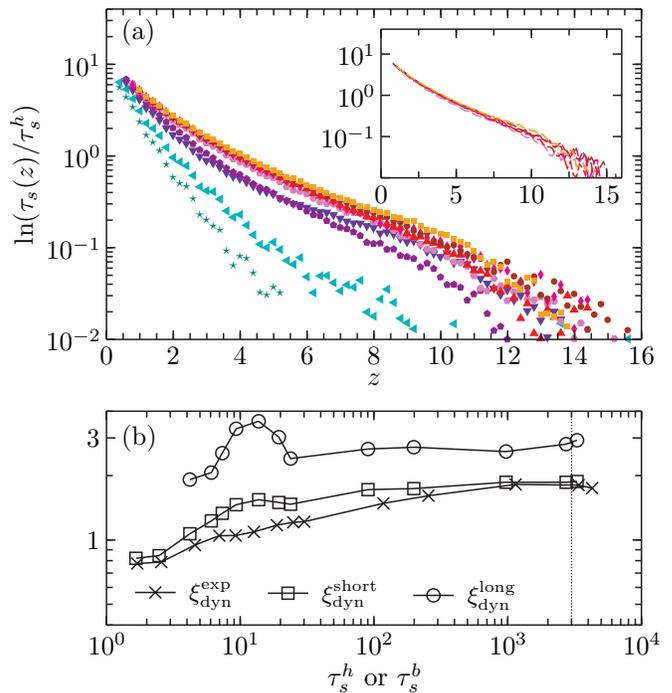}
\caption{(Color online) (a) Self-overlap relaxation times in the KA system divided by
value $\tau_s^h$, the value of $\tau_s(z)$ averaged over the large $z$ range $[15.4:15.8]$.
The temperatures shown her are the same as in Fig.~\ref{fig:tau_comparison}.
The relaxation is exponential at high $T$, becomes very curved at
intermediate temperature $T \approx 0.6$, and then evolves very little
at low $T$. For all temperatures, there is an exponential decay at
short distances, yielding a length $\xi^{\rm short}_{\rm dyn}$. There is
also an approximately exponential decay at long distances from which we
extract $\xi^{\rm long}_{\rm dyn}$. The inset isolates temperatures $0.432
\leq T \leq 0.5$ to emphasize the low-$T$ saturation. (b) Relaxation time
dependence of the three dynamic lengths extracted from fitting the profiles plotted
against $\tau_s^b$ for $\xi_{\rm dyn}^{\rm exp}$ and $\tau_s^h$ for
$\xi^{\rm long}_{\rm dyn}$ and $\xi^{\rm short}_{\rm dyn}$. The approximate position of
$T_c$ is indicated by a vertical dotted line.
Lengths are shown for all temperatures listed in Sec.~\ref{sec:methods}.}
\label{fig:taus_scaled}
\end{figure}

As an alternative approach for analyzing the data, we divide the profiles
$\tau_s(z)$ by $\tau_s^h$, the value of the relaxation time measured at
large $z$. We show the result in Fig.~\ref{fig:taus_scaled}(a). There
are several features worthy of discussion when viewed from this
perspective. First, except for the highest temperatures, the curves are
at $z_{max}$ still decaying which shows that the accuracy of our data
allows us to see finite size effects in the dynamics even in the center of
the box.  Second, while the highest temperatures display exponential
decay of $\ln(\tau_s)$, the curves at the lowest temperatures are
somewhat bent in this log-lin representation and it is not clear how
to best analyze quantitatively this curvature. As a first attempt,
we have tried to describe these data using an exponential decay at
short $z$, followed by a second exponential decay at longer $z$. At
intermediate supercooling, particularly at $T=0.6$, deviations from a
single exponential decay are indeed quite pronounced. Third, the low $T$
curves nearly collapse (except perhaps at very large $z$ values). This is
illustrated in the inset of Fig.~\ref{fig:taus_scaled}(a) where the curves
from the lowest five temperatures are isolated, clearly showing that
these profiles are virtually identical over a large range of distances,
despite the fact that relaxation times change by nearly two decades over
that same temperature regime (see Fig.~\ref{fig:tau_comparison}).

Since in this representation the data shows a clear curvature, we
have chosen to split it into two parts, one for $z\leq z^*$ and one
for $z>z^*$. From these fits we have extracted two dynamical length
scales, $\xi_{\rm dyn}^{\rm short}$ and $\xi_{\rm dyn}^{\rm long}$,
respectively. For $T>0.6$ best fits were obtained with $z^*=5$ and for
$T\leq0.6$ best fits resulted when using $z^*=6$. These lengths are shown
in Fig.~\ref{fig:taus_scaled}(b), along with the previous $\xi_{\rm
dyn}^{\rm exp}$, for reference. We have also fit these data with a sum of
exponentials, which provides a fit that matches the raw data quite well.
Unfortunately the large number of fit parameters leads to overfitting that results in nonphysical length scales.

It is important to emphasize that even with this refined method of
analysis, our previous results concerning the saturation of $\xi_{\rm
dyn}^{\rm exp}$ remain robust. Clearly, even without any fitting, there
is still a saturation of the dynamic profiles at temperatures
close to $T_c$, as can be seen from the collapse of the data with the new
normalization at low $T$ (see inset of Fig.~\ref{fig:taus_scaled}(a)). This
shows that the observation that $\xi_{\rm dyn}^{\rm exp}$ saturates as
a function of $T$ does not depend on the detail how the length scale
has been determined.

This more precise type of analysis permits to detect an unexpected
feature in the $T-$dependence of the dynamic length scales.
From Fig.~\ref{fig:taus_scaled}(b) we see that the absolute value and
the $T-$dependence of $\xi^{\rm short}_{\rm dyn}$ are very similar
to the ones of $\xi_{\rm dyn}^{\rm exp}$ in that also this length
scale grows with decreasing $T$ and then saturates around $T_c$. The
$T-$dependence of $\xi^{\rm long}_{\rm dyn}$ is somewhat weaker than
the one of $\xi^{\rm short}_{\rm dyn}$ in that it increases only
by a factor of around 1.5 instead of the factor of two found for
the latter. Much more important is, however, the observation that
at intermediate temperatures, $T\approx 0.6$, $\xi^{\rm long}_{\rm
dyn}$ shows a very pronounced peak. 
We stress that this peak is in no way a result of the fit and the behavior 
is clearly evident from inspection of the data in Fig.~\ref{fig:taus_scaled}(a).

We note that this temperature is
significantly above the mode-coupling temperature of the system, $T_c
\sim 0.435$, but it coincides with the crossover temperature recently
termed  ``$T_s$'' in Ref.~\onlinecite{Flenner-PRL2014}. The authors of
Ref.~\onlinecite{Flenner-PRL2014} identify this $T_s$ both with the breakdown
of the Stokes-Einstein relation as well as a change in the ``shape''
of dynamical heterogeneities. While a quantitative connection between
$\xi_{\rm dyn}^{\rm long}$ and the Stokes-Einstein decoupling is {\it a priori}
not obvious, we will in the following provide evidence that both types
of behavior are indeed correlated.

We emphasize that the non-monotonic evolution of the length
$\xi_{\rm dyn}^{\rm long}$ is a qualitatively different
phenomenon from the evolution of a dynamic length discussed in
Ref.~\onlinecite{Kob-NatPhys2012}. Indeed, because of the single exponential
fitting protocol used in that paper, the analog crossover at temperature
$T_s$ was previously not detected in the HARM system. In fact, while
$\ln(\tau_s)$ seems more exponential in the HARM system than in
the KA system, we indeed see a signature of this higher temperature
crossover around $T \approx 9$, near the putative $T_s$ for this
model, a temperature not examined in Ref.~\onlinecite{Kob-NatPhys2012} (see Appendix \ref{sec:taus_other}). In
particular, it should be noted that fits to obtain $\xi_{\rm dyn}^{\rm
exp}$ as shown in Fig.~\ref{fig:taus_scaled_exponential} and
in Ref.~\onlinecite{Kob-NatPhys2012} give relatively little weight to
the relaxation times at large $z$, thus making the peak seen in
$\xi^{\rm long}_{\rm dyn}$ is hardly noticeable in $\xi_{\rm dyn}^{\rm
exp}$  (see Fig.~\ref{fig:taus_scaled}).  Note that the discovery of
this novel crossover phenomenon does not affect any of the conclusions
about the dynamical length scale having a maximum at $T_c$ in this
system because we find, as in the KA system, that $\xi_{\rm dyn}^{\rm
exp} \approx \xi_{\rm dyn}^{\rm short}$, and both these length scales
have indeed a maximum in the vicinity of the mode-coupling crossover
temperature $T_c$.

To give further evidence for a relationship between the crossover
temperature $T_s$ and the evolution of the length $\xi_{\rm dyn}^{\rm
long}$, we have repeated our full analysis of dynamic profiles for a third
model, namely the WCA system, which again shows this anomalous behavior
around $T_s \approx 0.425$, as shown in Appendix \ref{sec:taus_other}.

\section{Comparison of transverse and longitudinal relaxation}
\label{sec:perppar}

In this section, we investigate the connection between dynamic
profiles and the crossover temperature $T_s$ further. Because
$T_s$ was previously related to a change in the geometry of dynamic
heterogeneities~\cite{Flenner-PRL2014}, a possible connection could come
from a geometric analysis of dynamic profiles near amorphous walls. To
this end, we separately analyze the relaxation into perpendicular and
parallel components with respect to the wall.

We analyze the self-intermediate scattering function at different
distances from the wall, as defined in Eq.~(\ref{eq:fsk}). It is simple
to resolve the relaxation times for dynamics parallel and perpendicular
to the wall, as done for instance in Ref.~\onlinecite{Scheidler-JCPB2004}. The
relaxation times obtained for wavevectors parallel to the wall are shown
in Fig.~\ref{fig:tau_comparison}. Here, $\tau^{||}(z)$ has been divided by
a constant value corresponding to the ratio $\tau^{||}(z)/\tau_s(z)$ at a
single distance ($z\approx15$) and a single temperature ($T=0.435$). The
curves and points respectively representing $\tau^{||}(z)$ and $\tau_s(z)$
lie perfectly on top of each other except for a small deviation
at the highest temperatures, suggesting that $\tau^{||}(z)$
reports on the same physics as $\tau_s(z)$ at supercooled temperatures,
and does not contain additional physical information about relaxation
near the amorphous wall.

\begin{figure*}
\centering
\includegraphics[scale=.95]{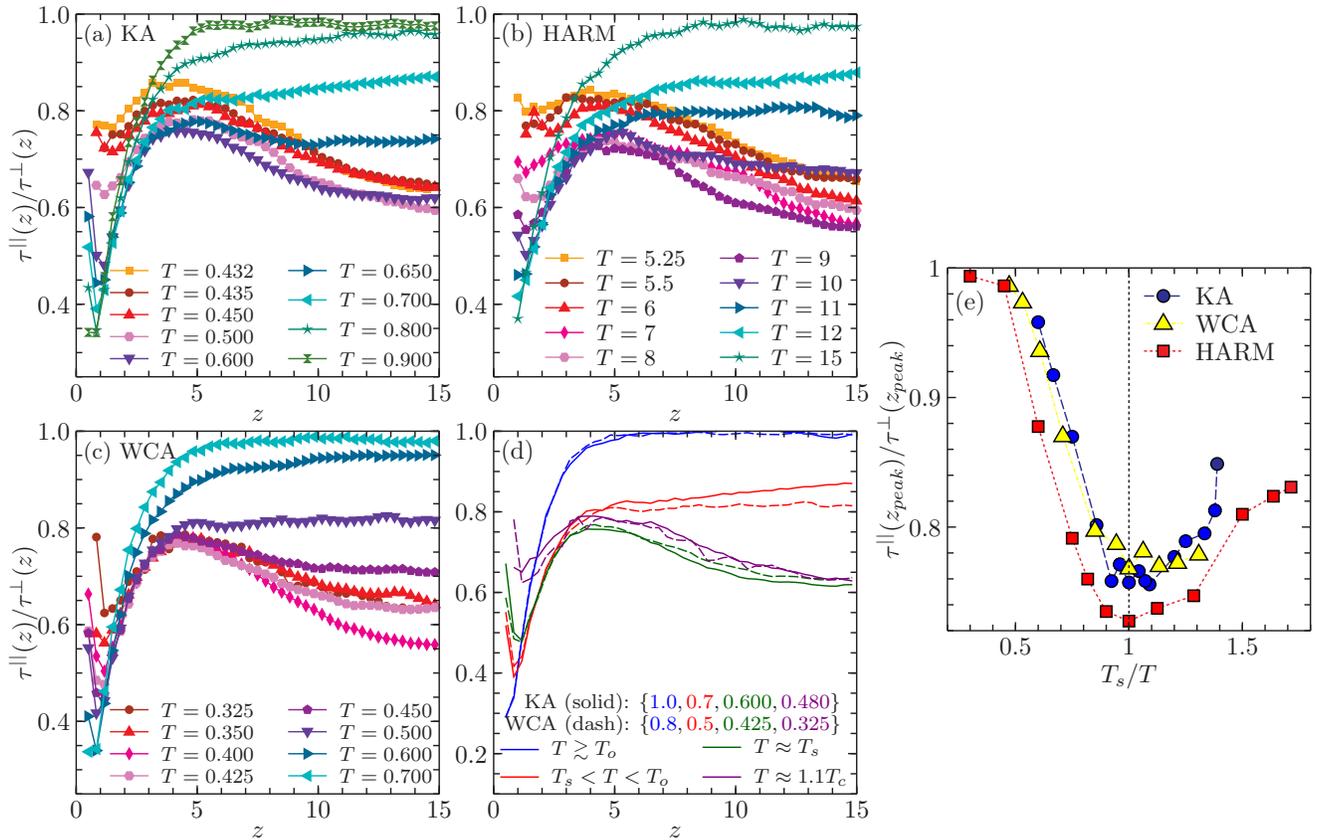}
\caption{(Color online) (a-c) Distance dependence of the ratio of the parallel to perpendicular
relaxation times in the three models studied. Due to the presence of
the wall, $\tau^{||}(z) \leq \tau^{\perp}(z)$ at all distances. At
high temperatures, the curves tends to unity but for lower temperatures the 
curves approach a fractional value for the $z$-range accessible.
 (d) Profiles of the curves in the
distinct regimes identified, emphasizing in particular the emergence
of a maximum in the ratio at a distance $z_{\rm peak}$. Results for KA are shown with solid lines
and for WCA with dashed lines. At $z=15$, from top to bottom the 
curves are ordered $T\gtrsim T_o$, $T_s < T < T_o$, $T\approx1.1T_c$, and $T\approx T_s$.
(e) The ratio 
of relaxation times at $z_{\rm peak}$ as a function of $T_s/T$,
showing that the peak value is minimized when $T \approx T_s$, the
crossover temperature analyzed in Ref.~\onlinecite{Flenner-PRL2014}. The
values used for $T_s$ are 0.6, 9.0, and 0.425 for the KA, HARM and WCA
systems, respectively.}
\label{fig:tau_ratio}
\end{figure*}

In agreement with previous results~\cite{Scheidler-JCPB2004},
the relaxation times obtained for wavevectors perpendicular to
the wall are always slower than for those parallel to the wall. In
Fig.~\ref{fig:tau_ratio}(a-c) we show how the perpendicular relaxation
times compare to the parallel relaxation times at different distances from
the wall via the investigation of the ratio $\tau^{||}(z)/\tau^\perp(z)$
in the three different models studied in this work, the HARM, KA and WCA
systems.

In each system three temperature regimes can be identified. The
qualitative features of these regimes are illustrated by
Fig.~\ref{fig:tau_ratio}(d), where we select a subset of equivalent
temperatures for each system, as being representative of the various
regimes, which can be described as follows.

\begin{itemize}

\item  Above $T_o$, the wall suppresses perpendicular more than parallel
relaxation at small $z$, but then the ratio tends rapidly towards unity
at large distances.

\item Near $T_o$, the ratio tends to a pseudo-plateau
at values less than one for all $z$ values accessible. A dip in the
ratio of $\tau^{||}(z)/\tau^{\perp}(z)$ is also observed to develop near
$z\approx1$. This results in a maximum of the ratio at an intermediate
distance $z_{\rm peak} \approx 4.0$ as $T$ is lowered.

\item At temperatures below $T_o$, the curves become
more complex in that at large $z$ the values of the ratio decreases
whereas at $z_{\rm peak}$ the ratio is at intermediate $T$ basically
constant before it increases slightly. Furthermore we see that the dip
at $z\approx 1$ also rises slightly with decreasing temperature.

\end{itemize}

We define $z_{\rm peak}$ as the distance from the wall where the ratio
$\tau^{||}/\tau^\perp$ peaks and consider the temperature evolution of
this peak value, as shown in Fig.~\ref{fig:tau_ratio}(e). In all three
systems, the temperature where the peak value is minimized falls very
close to values of $T_s$ given in Ref.~\onlinecite{Flenner-PRL2014}.

This coincidence is robust against finite size effects that are
known to affect the dynamics in simulation boxes with an elongated
shape~\cite{Coslovich-2014}. Tests using larger box sizes demonstrate
that the precise value of the ratio at large $z$ depends weakly
on the box size, but the data up to $z_{\rm peak}$ and just beyond
are fairly insensitive to such changes, as demonstrated in Appendix
\ref{sec:fse}. Hence the behavior and temperature correspondence appear
to be fairly insensitive to finite size effects.

The precise dependence of the ratio $\tau^{||}(z)/\tau^\perp(z)$
is complicated, and further microscopic investigations of
its behavior would be needed to understand these data in more
detail. However, it is interesting to speculate on the connection
between the behavior observed in Fig.~\ref{fig:tau_ratio}(e)
and that reported in Ref.~\onlinecite{Flenner-PRL2014}. The authors
of  Ref.~\onlinecite{Flenner-PRL2014} note that the ratio of dynamical
heterogeneity length scales associated with parallel and perpendicular
displacements markedly changes behavior at $T_s$, which they claim to be
the temperature at which violation of the Stokes-Einstein relation first
occurs. The fact that we observe in the vicinity of $T_s$ a significant
change in the $T-$dependence of $\tau^{||}(z)/\tau^\perp(z)$ is
harmonious with the notion that this temperature is associated with a
change in directionally resolved relaxation motifs, and gives further
impetus for microscopic study of particle motion near the frozen wall
as $T$ varies above and below $T_s$.

It should also be noted that we observe clear features
of altered relaxation in both $\xi_{\rm dyn}^{\rm long}$ as well as
$\tau^{||}(z)/\tau^\perp(z)$ at $T_s$, independent of metrics based
on Stokes-Einstein violation, whose onset is not necessarily sharp
enough to clearly define a characteristic temperature. In this sense,
one can bypass definitions based on transport anomalies and relate $T_s$
directly to the change of two-point relaxation behavior provided by the
proximity of a frozen interface.

In summary we can conclude that the results of this section and
Sec.~\ref{sec:overlap} above give further evidence in support of the
notion, first advanced in Ref.~\onlinecite{Flenner-PRL2014}, of a well-defined
characteristic temperature $T_s$ below the onset temperature of slow
dynamics but above $T_c$, which physically relates to a marked crossover
in the geometric properties of dynamic heterogeneity.

\section{Discussion and Conclusions}
\label{sec:conclusions}

In this work we have extended previous studies related to the relaxation of
supercooled liquids near a wall created from a subset of particles
fixed in their equilibrium positions. We find clear evidence
that as $T_c$ is approached, the dynamical length scale defined
%note in next line citation split up and in reverse order because this matches current citation order
in Refs.~\onlinecite{Kob-NatPhys2012} and \onlinecite{Scheidler-EPL2002} quantitatively
saturates in the KA system. This behavior is to be contrasted with the
non-monotonic growth of the same length scale in the HARM system. This
distinction, namely a saturation as opposed to a decrease in the
length scale below $T_c$, is in harmony with both the behavior of
$\chi_4^{NVE}$~\cite{Berthier-JCP2007,Flenner-JCP2013} as well as trends
in the behavior of finite size effects as filtered through relaxation
times in these two models~\cite{Berthier-finite-PRE2012}. Taken together,
these results all suggest that the crossover between transport mechanisms
at $T_c$ is qualitatively similar in both systems, but is quantitatively
sharper in the HARM, which may therefore be viewed to be closer to
the idealized mean-field limit than the KA system.

We have also investigated the behavior of relaxation channels parallel
and perpendicular to the wall. While motion parallel to the amorphous
wall mirrors the behavior revealed by studies of the self-overlap function,
we found that the behavior of the ratio $\tau^{||}(z)/\tau^\perp(z)$
shows clear evidence of a change in behavior at a recently identified
temperature $T_s$, across three different model systems. In
Ref.~\onlinecite{Flenner-PRL2014} it has been argued that $T_s$ marks a
temperature where the shape of the dynamical heterogeneities changes, in
the sense that transverse and longitudinal relative motions of particles
become decoupled. Thus it is physically reasonable that close to the
wall such a decoupling also affects the relaxation behavior in the
parallel direction in a different manner than in the orthogonal one,
thus rationalizing our findings regarding the $T-$ and $z-$dependence
of $\tau^{||}(z)/\tau^\perp(z)$.

While our results place the change of transport mechanisms invoked to account for the 
dynamics of the HARM model \cite{Kob-NatPhys2012,Kob-PP2012} 
near an amorphous wall on firmer grounds, we emphasize that the strong 
saturation of the dynamic lengthscale $\xi_{\rm dyn}^{\rm exp}$ revealed 
in both HARM and KA models has no clear counterpart in available
measurements of dynamic lengthscales from bulk four-point functions \cite{Flenner-Comment-NatPhys2012,Kob-Reply-NatPhys2012},
which appear to display no obvious saturation in the mode-coupling regime.
Although this might indicate that both types of measurements are unrelated, we also 
note that the crossover temperature $T_s$ detected through analysis of 
four-point functions in Ref.~\onlinecite{Flenner-PRL2014} is also observed here 
using measurements near an amorphous wall via two-point
quantities, albeit those extracted near an object that breaks spatial symmetry.
While this approach is in the same spirit of earlier studies based on
measurements of the response of two-point correlators to external
fields \cite{Berthier-JCP2007,berthier-science}, the present set-up
using pinned particles provides a potentially simple means to observe
the changes that occur at $T_s$ in an experimental setting. Future work
should be devoted to real-space dynamical analysis of the behavior of
motion parallel and perpendicular to the wall in order to gain deeper
insight into the observations presented in this work.
Overall, these results reveal that a better understanding of the 
connection between the various dynamic lengthscales studied 
in supercooled liquids is needed.

\begin{acknowledgments}
We thank D. Coslovich, E. Flenner, and G. Szamel for useful
discussions. Some of this research was performed on resources
provided by the Extreme Science and Engineering Discovery Environment
(XSEDE), which is supported by National Science Foundation (NSF)
grant number OCI-1053575. Some computations were performed on the
Midway resource at the University of Chicago Research Computing Center
(RCC). Simulations performed were organized by executing LAMMPS runs
with the Swift parallel scripting language (NSF Grant No.~OCI-1148443)
\cite{Wilde-Parallel2011}. G. M. H and D. R. R. were supported by NSF
grants DGE-07-07425 and CHE-1213247, respectively. W.K. acknowledges support
from the Institut Universitaire de France. The research leading
to these results has received funding from the European Research Council
under the European Union's Seventh Framework Programme (FP7/2007-2013)
/ ERC Grant agreement No 306845.  
\end{acknowledgments}

\begin{appendix}

\section{Static Overlaps}
\label{sec:static}

\begin{figure}
\centering
\includegraphics{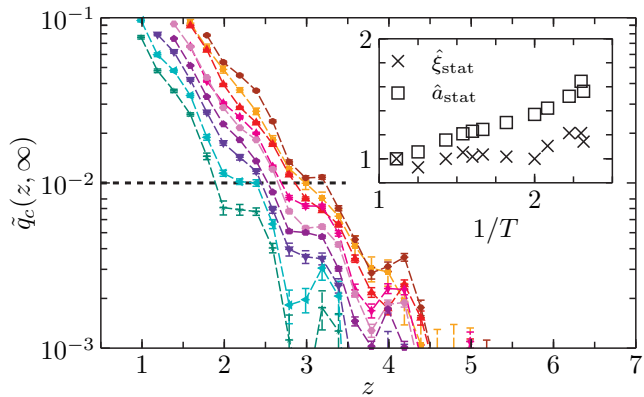}
\caption{(Color online) Static overlap values extracted from $q_c(z,t \to \infty)$ for the
temperatures in Fig.~\ref{fig:tau_comparison}. The
decay is exponential and becomes steadily longer ranged as temperature is
lowered. Two static length scales, $\xi_{\rm stat}$, the inverse of the slope
of the decay, and $a_{\rm stat}$, the $z$ value where $\tilde{q}_c(z,\infty)$
lines cross the value $C=0.01$, as illustrated by a horizontal dashed line
may be extracted. $1\sigma$ error bars from a bootstrap analysis are shown
\cite{Efron1993-Bootstrap}.
Inset: The lengths $\hat{a}_{\rm stat}=a_{\rm stat}/a_{\rm stat}^0$
and $\hat{\xi}_{\rm stat}=\xi_{\rm stat}/\xi_{\rm stat}^0$ where $a_{\rm stat}^0$=1.92
and $\xi_{\rm stat}^0$=0.56 are the values of the lengths at high temperature,
$T=0.9$.}
\label{fig:qinf}
\end{figure}

Although the focus of this work is on the dynamical properties of the
KA system near an amorphous wall, it is also useful to investigate the
growth of static order. In Fig.~\ref{fig:qinf} we show the static overlaps
for comparison with Ref.~\onlinecite{Kob-NatPhys2012}. As in that work, we
see that the static overlap decays exponentially with distance from the wall,
and we remark that these static quantities decay much more quickly
than the dynamical profiles shown in Figs.~\ref{fig:taus_scaled}. We
see a slight ``layering'' effect which we suspect is more pronounced
here than in Ref.~\onlinecite{Kob-NatPhys2012} because in the present work we
used a completely amorphous wall, while in Ref.~\onlinecite{Kob-NatPhys2012}
a reflective wall and an amorphous potential were used in combination.

To extract static length scales, we choose to fit these curves to the function 
\begin{equation}
\tilde{q}_c(z,\infty) = C \exp(-(z-a_{\rm stat})/\xi_{\rm stat}),
\end{equation} 

\noindent
with a fixed value of $C=0.01$, shown by a horizontal dashed line
in Fig.~\ref{fig:qinf}. This fit gives values for $\xi_{\rm stat}$
describing the exponential decay of the static profiles, which are
small and grow very slowly. In addition, we can consider the evolution
of $a_{\rm stat}$ quantifying the distance $z$ for which the overlap equals
$C=0.01$. This is another static length scale, which is slightly larger
than $\xi_{\rm stat}$ and also grows quite slowly.

Both are shown in the inset of Fig.~\ref{fig:qinf} and are seen
to grow by a much smaller amount than the dynamical length scales
measured. Although there are small qualitative differences between the KA
and HARM systems with respects to the magnitudes of static and dynamic
length scales, both show a pronounced decoupling between static and
dynamic properties, in line with other studies showing static length
scales which are smaller and grow more slowly than dynamical ones
\cite{Karmakar-PNAS2009,Berthier-static-PRE2012,Hocky-PRL2012,Charbonneau-PRE2013}.

\section{Relaxation times in the KA system}
\label{sec:tauT}
\begin{figure}[ht]
\centering
\includegraphics[scale=0.9]{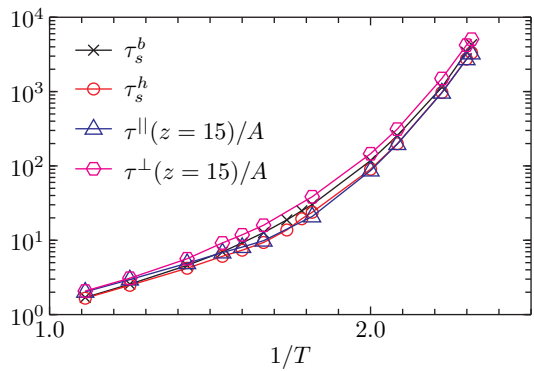}
\caption{(Color online) Temperature dependence of the various relaxation times
discussed in the main text for the KA system. The values of $\tau^{||}$ and $\tau^\perp$ 
have been scaled by the value $A=0.7$ as discussed in Fig.~\ref{fig:tau_comparison}.  }
\label{fig:tau_invt}
\end{figure}

Fig.~\ref{fig:tau_invt} compares the four relaxation times used in the main text for the KA system. As discussed previously, the values of $\tau^{||}(z)$ and $\tau_s(z)$ can be scaled on top of each other using a temperature independent constant, although there is some deviation at high temperatures. The need for rescaling simply arises because of the choice of the window size $l=0.45$ for the self-overlap calculation and the value $k=7.25$ for the self-intermediate scattering function. In Fig.~\ref{fig:tau_invt}, we have also rescaled the values of $\tau^\perp$ showing that relaxation in the perpendicular direction in the middle of the box is slower than in the parallel direction at all temperatures. Finally, we also show the value of the self-overlap relaxation time without the wall, which is larger than the self-overlap relaxation time at all temperatures at the center of the box in the presence of the wall. This origin of this result will be discussed in a forthcoming work \cite{Coslovich-2014}.

\section{Overlap curves in the HARM and WCA systems}

\begin{figure}[ht]
\centering
\includegraphics{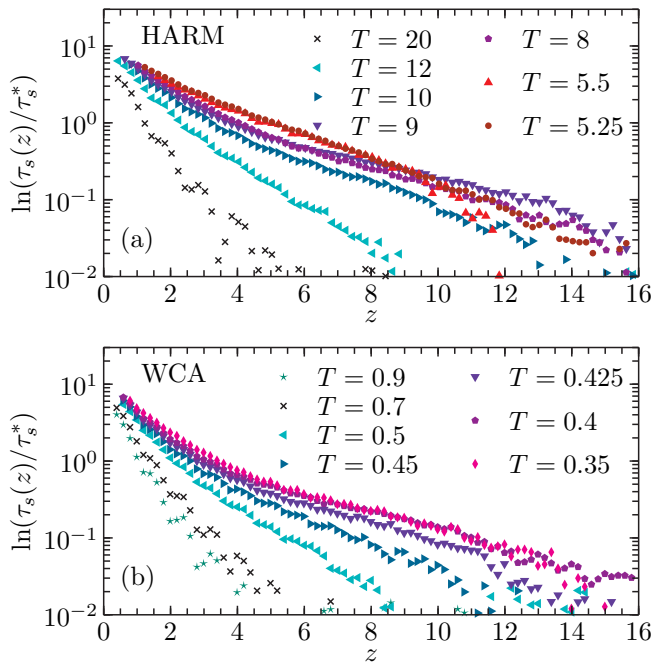}
\caption{(Color online) Distance dependence of the self-overlap relaxation times for
the HARM system (a) and the WCA system (b) in a representation
similar to the one of Fig.~\ref{fig:taus_scaled} for the KA model.}
\label{fig:taus_other}
\end{figure}

For the sake of comparison with the results presented in
Fig.~\ref{fig:taus_scaled}, we also include the results for the HARM
and WCA systems. In Fig.~\ref{fig:taus_other}(a) the results for the HARM
system are plotted. While the curves decay in a completely exponential
manner at the highest and lowest temperatures, there is a pronounced
curvature or double-exponential character near $T=9$, a temperature
not studied in Ref.~\onlinecite{Kob-NatPhys2012}. The curvature at $T=8$
is consistent with the data in that work.

Figure \ref{fig:taus_other}(b) shows results for the WCA system.  Here
again the data appears to have double exponential character starting
near $T=0.425$. We note for both models that this behavior occurs in
the range identified as $T_s$ by Ref.~\onlinecite{Flenner-PRL2014}. A more
extensive study should test whether there is any difference using
a smoothed WCA potential, which has slightly different dynamical
properties at low temperatures from those of the standard WCA (see
e.g. Ref.~\onlinecite{Coslovich-PRE2011}).

\label{sec:taus_other}

\section{Finite Size Effects}
\label{sec:fse}

\begin{figure}[ht]
\centering
\includegraphics{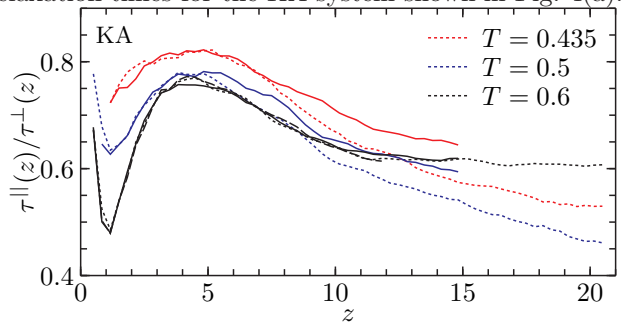}
\caption{(Color online) Ratio of parallel to perpendicular relaxation times in three
different box sizes. The size used in the main text is indicated by a
solid line, while data for a larger box size is shown with dotted lines,
and for a smaller box at $T=0.6$ by a dashed line, showing that finite
size effects do not affect the crossover behavior at $T=T_s$ discussed
in Sec.~\ref{sec:perppar}.}
\label{fig:tau_ratio_fse}
\end{figure}

We have performed additional simulations to check that our conclusions
are robust with respect to changes in system sizes. As an example of such
tests, we consider the behavior of the ratio of perpendicular and parallel
relaxation times for the KA system shown in Fig.~\ref{fig:tau_ratio}(a).

In Fig.~\ref{fig:tau_ratio_fse} we show the same calculation for the KA
system using $N=7600$ in a box with aspect ratio $1:1:4$. We also show
at $T=T_s=0.6$ the behavior for the $N=5700$ system in an effectively
smaller box, generated by freezing a wall of thickness $W=9$ rather than
$W=3$ as in the main body of this article. While the large-$z$ behavior
seems to depend on the system size, the data up to $z_{\rm peak}$ appear to
be independent of the box size considered. Hence we conclude that our
approach to analyze the data in Fig.~\ref{fig:tau_ratio}(e) to define $T_s$
is safe, and that the existence of a correspondence between the crossover
$T_s$ of Ref. \onlinecite{Flenner-PRL2014} and the present comparative study
of parallel and perpendicular relaxation times is a robust finding.

Finally, we note that the large-$z$ values of these data change with system size. We believe that this is due to complex hydrodynamic coupling between the dynamics in the two directions. In fact, we found {\em in the bulk} that for $T<T_o$ in a non-cubic box, the ratio of $\tau^{||}(z)/\tau^{\perp}(z)$ is non-unity due to hydrodynamic effects (data not shown). For example, in the KA system, this ratio is approximately 0.85 for a $3:1:1$ box at temperatures $T\leq0.8$. Study of the precise origin of this phenomenon is beyond the scope of this work and will require further investigation.

\end{appendix}

%\clearpage

\bibliography{kalj_wall}

\end{document}